\documentstyle[11pt]{article}

\begin{document}

\begin{center}
{\huge {\bf Probability distribution of arrival times in quantum mechanics 
\vspace*{0.6cm}\\}} {\Large {V. Delgado \vspace*{.2cm}}}\\{\it Departamento
de F\'\i sica Fundamental y Experimental, \\Universidad de La Laguna,
38205-La Laguna, Tenerife, Spain\vspace*{0.7cm}\\}
\end{center}

\begin{abstract}
In a previous paper [V. Delgado and J. G. Muga, Phys. Rev. A {\bf 56}, 3425 
(1997)] we introduced a self-adjoint 
operator $\hat {{\cal T}}(X)$ whose eigenstates can be used to define
consistently a probability distribution of the time of arrival at a given
spatial point. In the present work we show that the probability distribution
previously proposed can be well understood on classical grounds in the sense
that it is given by the expectation value of a certain positive definite
operator $\hat J^{(+)}(X)$ which is nothing but a straightforward quantum
version of the modulus of the classical current. For quantum states highly
localized in momentum space about a certain momentum $p_0 \neq 0$, the
expectation value of $\hat J^{(+)}(X)$ becomes indistinguishable from the
quantum probability current. This fact may provide a justification for the
common practice of using the latter quantity as a probability distribution
of arrival times.
\end{abstract}

\vspace{1.8 cm}

\begin{center}
{\large {\bf {I. INTRODUCTION\vspace{.2 cm}\\}}}
\end{center}

Standard quantum mechanics is mainly concerned with probability
distributions of measurable quantities at a given instant of time. Such
distributions can be inferred from the formalism in terms of projections of
the instantaneous state vector $|\psi (t)\rangle $ onto appropriate
subspaces of the whole Hilbert space of physical states. However, one may
also be interested in the probability that a certain physical quantity takes
a definite value between the instants of time $t$ and $t+dt$. Let us assume
this quantity to be the position of a particle and restrict ourselves to one
spatial dimension. What is the probability distribution of arrival times at
a detector situated at a given point $x=X$? Standard quantum theory is based
on the assumption that, by reducing sufficiently the experimental
uncertainties involved, the outcomes of any measurement process will
reproduce, within any desirable precision, the ideal distribution inferred
from the spectral decomposition of a certain self-adjoint operator 
associated with the physical quantity under consideration. It is therefore 
implicitly assumed that the quantum formalism can provide a prediction for
the experimental results without having to make reference to the specific
properties of the measuring device involved. Since the distribution of
arrival times at a given spatial point is, in principle, a measurable
quantity that can be determined via a time-of-flight experiment, it is
reasonable to ask for an apparatus-independent theoretical prediction.

In classical statistical mechanics the above question has a definite answer:
The (unnormalized) probability distribution of arrival times at $X$ for a
certain statistical ensemble of particles of mass $m$, moving along a
well-defined spatial direction (i.e., either with momenta $p>0$ or with
momenta $p<0$), is given by the average current at $X$,

\begin{equation}
\label{jota}\langle J(X)\rangle =\!\int \!\!\int \!f(x,p,t)\,\frac
pm\,\delta (x-X)\,dx\,dp, 
\end{equation}
where $f(x,p,t)$ represents the phase-space distribution function
characterizing the statistical ensemble. In quantum mechanics, however,
things turn out to be much more involved. In particular, a straightforward
application of the correspondence principle would lead us to consider the
expectation value of the current operator

\begin{equation}
\label{curop}\hat J(X)=\frac 1{2m}\left( \hat P\,|X\rangle \langle
X|+|X\rangle \langle X|\,\hat P\right) 
\end{equation}
($\hat P$ denoting the momentum operator) as the most natural quantum
candidate for the probability distribution of the time of arrival at a point 
$X$. However, even though such a definition has been widely used in recent
times [\ref{Du}--\ref{Mug}], it cannot be considered as a satisfactory
solution because of the fact that the expectation value of $\hat J(X)$ is
not positive definite, even for wave packets containing only
positive-momentum components. Nonetheless, when quantum backflow
contributions become negligible, one expects the expectation value of the
current operator to be a good approximation to the actual probability
distribution of arrival times.

The difficulty for defining such probability distributions is nothing but a
mere aspect of the more fundamental problem of the nonexistence of a quantum
time operator conjugate to the Hamiltonian. The reason for this latter fact
lies, basically, in the incompatibility of such a time operator with the
semibounded nature of the Hamiltonian spectrum [\ref{Pauli}--\ref{VDB4}].

In spite of detailed work by Allcock [\ref{Allco}] denying the possibility
of incorporating the time-of-arrival concept in the quantum framework, more
recently there has been considerable effort in defining a probability
distribution of the time of arrival of a quantum particle at a given spatial
point [\ref{Du}--\ref{Mug}, \ref{VDB4}--\ref{Gian}]. The incorporation of
such probability distributions in the formalism of quantum mechanics has
both conceptual and practical interest. In particular, this issue is closely
related to the problem of the temporal characterization of tunneling (the 
so-called tunneling time problem) [\ref{Lan}--\ref{VDB2}], whose 
understanding is important for its possible application in semiconductor 
technology.

In Ref. [\ref{Du}], Dumont and Marchioro, primarily concerned with
tunneling-time distributions, proposed the probability current as a quantum
definition of the (unnormalized) probability distribution of arrival times
at a point sufficiently far to the right of a one-dimensional potential
barrier. Leavens [\ref{Leav2}] has shown that this result can also be
derived within Bohm's trajectory interpretation of quantum mechanics by 
making the assumption that particles are not reflected back through the 
point $X$ (i.e., $\langle \psi (t)|\hat J(X)|\psi(t)
\rangle < 0$ does not occur for any $t$). On the other hand, Muga {\em et 
al. }[\ref{Mug}]{\em \ } have provided an operational justification of such 
a definition by simulating the detection of incoming particles by a 
destructive procedure.
More recently, Grot {\em et al. }[\ref{Grot}]{\em \ }have faced the problem
from a somewhat different perspective. These authors construct a suitable
self-adjoint operator in order to infer a probability distribution of
arrival times from its spectral decomposition. More specifically, starting
from the classical equations of motion for a particle moving freely in one
spatial dimension and solving for the time, they arrive at the operator

\begin{equation}
\label{trove1}\hat T(X)=\sqrt{\frac m{\hat P_{{\rm H}}(0)}}\,\left[ X-\hat
X_{{\rm H}}(0)\right] \,\sqrt{\frac m{\hat P_{{\rm H}}(0)}} 
\end{equation}
as a natural candidate for the time of arrival of a quantum free particle at
the spatial point $X$. In the above equation $\hat X_{{\rm H}}(t)$ and $\hat
P_{{\rm H}}(t)$ denote the position and momentum operators in the Heisenberg
picture, respectively, and are related to the corresponding Schr\"odinger
operators by

\begin{equation}
\label{heisc}\hat O_{{\rm H}}(t)=e^{i\hat H_0t/\hbar }\,\hat O\,e^{-i\hat
H_0t/\hbar }, 
\end{equation}
where $\hat O$ stands for $\hat X$ or $\hat P$, and $\hat H_0=\hat P^2/2m$
is the Hamiltonian of the free particle. The operator (\ref{trove1}) has the
interest that it represents a quantum version of the corresponding classical
expression $t(X)=m[X-x(0)]/p(0)$ obtained by straightforward application
of the correspondence principle and the {\em canonical quantization method} 
[\ref{QF}], which states that classical equations remain formally valid in
the quantum framework provided that one makes the substitution of Poisson
brackets by commutators $\{\;,\;\}\rightarrow 1/i\hbar \,[\;,\;]$ and
interprets the classical dynamical variables as self-adjoint operators in
the Heisenberg picture. Of course, whenever the classical expression under
consideration contains products of dynamical variables having a nonvanishing
Poisson bracket (as is the case for position and momentum), the mere
application of the canonical quantization method does not guarantee the
unambiguous construction of the corresponding quantum quantity. In fact, a 
specific symmetrization or quantization rule has been explicitly chosen in
Eq. (\ref{trove1}). For instance, another possible quantum operator obtained
from the same classical expression via a different symmetrization rule
(which has been previously introduced by Aharonov and Bohm [\ref{Aharo}]) is

\begin{equation}
\label{tahar1}\hat T(X)=\frac 12\left( \left[ X-\hat X_{{\rm H}}(0)\right]
\frac m{\hat P_{{\rm H}}(0)\,}+\frac m{\hat P_{{\rm H}}(0)\,}\,\left[ X-\hat
X_{{\rm H}}(0)\right] \right) . 
\end{equation}

Unfortunately, despite the fact that the above two operators are exactly
what one would expect by virtue of the correspondence principle, none of
them is self-adjoint. To circumvent this difficulty, Grot {\em et al.}
proposed a modified time operator such that, when acting on states with no
zero-momentum components, it leads to the same results as the operator (\ref
{trove1}) defined above.

In a previous paper [\ref{VDB4}] we followed a different route: Guided by
the fact that, in general, a self-adjoint time operator conjugate to the
Hamiltonian does not exist, we instead looked for a self-adjoint operator 
$\hat {{\cal T}}(X)$ with dimensions of time, conjugate to a conveniently
defined energy operator having a nonbounded spectrum. We showed that the
orthogonal spectral decomposition of such an operator can be used to define
consistently a probability distribution of arrival times at a given spatial
point within the standard formalism of quantum mechanics.

In this paper we are mainly interested in relating the formulation proposed
in Ref. [\ref{VDB4}] with the corresponding classical formulation. A quantum
expression looks more natural as long as it is possible to derive it from a
known classical quantity by applying certain specific quantization rules
(even though such a procedure is by no means a necessary condition for the
validity of a quantum formulation). We begin by briefly reviewing the
relevant formalism in Sec. II. In Sec. III we consider the
semiclassical limit of the proposed probability distribution of arrival
times. In Sec. IV we show that such a probability distribution can be
well understood on classical grounds in the sense that it is formally
analogous to its corresponding classical counterpart. In this section we
also provide a relation between the expectation value of the self-adjoint
operator $\hat {{\cal T}}(X)$ and the expectation values of the operators 
$\hat T(X)$ given by Eqs. (\ref{trove1}) and (\ref{tahar1}). In Sec. V we
analyze under what circumstances the proposed probability distribution can 
be replaced, to a good approximation, by the probability current, which has
been frequently used, in practice, as a quantum probability distribution of
arrival times. Finally, we conclude in Sec. VI.

\vspace{1.8 cm}

\begin{center}
{\large {\bf {II. FORMALISM\vspace{.2 cm}\\}}}
\end{center}

In looking for a probability distribution of arrival times within the
framework of standard quantum mechanics, we introduced in Ref. [\ref{VDB4}]
a self-adjoint energy operator $\hat {{\cal H}}$ defined by

\begin{equation}
\label{seop}\hat {{\cal H}}\equiv {\rm sgn}(\hat P)\,\frac{\,\hat P^2}{\,2m}%
, 
\end{equation}
where, in terms of a basis of momentum eigenstates $\left\{ p\rangle
\right\} $, the operator ${\rm sgn}(\hat P)$ reads

\begin{equation}
\label{sgdf}{\rm sgn}(\hat P)\equiv \!\int_0^\infty dp\left( |p\rangle
\langle p|-\mid \!-p\rangle \langle -p\!\mid \right) . 
\end{equation}
The normalization has been chosen so that the states $|p\rangle $ satisfy
the closure and orthonormalization relations

\begin{equation}
\label{ec12}\int_{-\infty}^{+\infty} dp\,|p\rangle \langle p|={\bf 1}, 
\end{equation}

\begin{equation}
\label{ec13}\langle p|p^{\prime }\rangle =\delta (p-p^{\prime }). 
\end{equation}

The motivation for introducing the operator $\hat {{\cal H}}$, which
essentially represents the energy of the free particle with the sign of its
momentum, lies in the fact that, unlike the Hamiltonian, it exhibits a
nonbounded spectrum. It is therefore possible to define a self-adjoint
operator with dimensions of time $\hat {{\cal T}}(X)$ by simply demanding
it to be conjugate to $\hat {{\cal H}}$, i.e.,

\begin{equation}
\label{ec80}[\hat {{\cal H}},\hat {{\cal T}}(X)]=e^{-i\hat PX/\hbar }\,[\hat
{{\cal H}},\hat {{\cal T}}(0)]\,e^{+i\hat PX/\hbar }=i\hbar . 
\end{equation}
This procedure led us to an operator $\hat {{\cal T}}(X)$ whose orthogonal
spectral decomposition reads

\begin{equation}
\label{ec68}\hat {{\cal T}}(X)=\!\int_{-\infty }^{+\infty }\!d\tau \,\tau
\,|\tau ;X\rangle \langle \tau ;X|, 
\end{equation}
where the states $|\tau ;X\rangle $, which constitute a complete and
orthogonal set, are given by

\begin{equation}
\label{ec69}|\tau ;X\rangle =h^{-1/2}\!\int_{-\infty }^{+\infty }\!dp\sqrt{%
\frac{|p|}m}\,e^{i({\rm sgn}(p)\frac{p^2}{2m}\tau -\,pX)/\hbar }\,|p\rangle. 
\end{equation}

In order to facilitate an interpretation in terms of measurement results, it
turns out to be most convenient to decompose the eigenstates $|\tau
;X\rangle $ as a superposition of negative- and positive-momentum
contributions, in the form

\begin{equation}
\label{ec71}|\tau ;X\rangle =|t\!=\!-\tau ,-;X\rangle +|t\!=\!+\tau
,+;X\rangle , 
\end{equation}
with $|t,\pm ;X\rangle $ defined by

\begin{equation}
\label{tpmh}|t,\pm ;X\rangle =h^{-1/2}\!\int_0^\infty \!dp\sqrt{\frac pm}%
\,e^{i(\frac{p^2}{2m}t\mp \,pX)/\hbar }\mid \!\!\pm p\rangle . 
\end{equation}
As can be easily verified, even though the states $|t,\pm ;X\rangle $
constitute a complete set they are not orthogonal. Specifically,

\begin{equation}
\label{copt}\sum_{\alpha =\pm }\!\int_{-\infty }^{+\infty }\!dt\;|t,\alpha
;X\rangle \langle t,\alpha ;X|={\bf 1}, 
\end{equation}

\begin{equation}
\label{ec32}\langle t,\alpha |t^{\prime },\alpha ^{\prime }\rangle =\frac
12\delta _{\alpha \alpha ^{\prime }}\left\{ \delta (t-t^{\prime })-P\frac
i{\pi (t-t^{\prime })}\right\} . 
\end{equation}
Despite this fact, the decomposition (\ref{ec71}) is interesting because the
variable $t$, unlike $\tau $, admits a proper interpretation as a physical
time. In particular, the states $|t,\pm ;X\rangle $ not only have the
desirable time-translation property 
\begin{equation}
\label{ec30b}e^{i\hat H_0t^{\prime }/\hbar }\,|t,\pm ;X\rangle =|t+t^{\prime
},\pm ;X\rangle , 
\end{equation}
but also transform under time reversal as $\,|t,\pm \rangle \rightarrow
|-t,\mp \rangle $.

Consider a free particle propagating along the $x$ axis toward a detector
located at a given point $X$. We shall assume that its actual state at $t=0$
is, in the position representation, either a linear superposition of
positive plane waves $|\psi _{+}(0)\rangle $ (corresponding to particles
arriving at the detector from the left) or a linear superposition of
negative plane waves $|\psi _{-}(0)\rangle $ (corresponding to particles
arriving at the detector from the right). At any instant of time the state
vectors $|\psi _{\pm }(t)\rangle $ satisfy the identity

\begin{equation}
\label{psit}|\psi _{\pm }(t)\rangle \equiv \Theta (\pm \hat P)\,|\psi _{\pm
}(t)\rangle , 
\end{equation}
where $\Theta (\pm \hat P)$ are projectors onto the subspaces spanned by
plane waves with positive/negative momenta

\begin{equation}
\label{ec16}\Theta (\pm \hat P)=\!\int_0^\infty dp\mid \!\pm p\rangle
\langle \pm p\!\mid . 
\end{equation}
It can be shown that for normalizable states satisfying Eq. (\ref{psit}) and
vanishing (in momentum representation) faster than $p$ as $p\rightarrow 0$,
it holds that [\ref{VDB4}]

\begin{eqnarray}
\pm \langle \psi _{\pm }|\,\hat {{\cal T}}(X)\,|\psi _{\pm
}\rangle &=&\int_{-\infty }^{+\infty }d\tau \,\tau \,\langle \psi _{\pm }
\mid\!\pm \tau ;X\rangle \langle \pm \tau ;X\!\mid \psi _{\pm }\rangle 
\nonumber \\
\label{ec62} &=&\frac{\int_{-\infty }^{+\infty }d\tau \,\tau \,\langle 
\psi _{\pm }(\tau )|\hat
J(X)|\psi _{\pm }(\tau )\rangle }{\int_{-\infty }^{+\infty }d\tau \,\langle
\psi _{\pm }(\tau )|\hat J(X)|\psi _{\pm }(\tau )\rangle }, 
\end{eqnarray}
where use has been made of Eq. (\ref{ec68}) and $|\psi _{\pm }\rangle $
denotes the state of the particle in the Heisenberg picture,

\begin{equation}
\label{ec39}|\psi _{\pm }\rangle =e^{i\hat H_0\tau /\hbar }\,|\psi _{\pm
}(\tau )\rangle =|\psi _{\pm }(0)\rangle . 
\end{equation}

As already stated, the right-hand side of Eq. (\ref{ec62}) can be recognized
as a quantum version of the mean arrival time at $X$, obtained by
straightforward application of the correspondence principle to the analogous
classical expression for a statistical ensemble of particles propagating
along a well-defined spatial direction. Furthermore, the positive-definite
quantity $\langle \psi _{\pm }\mid \!\pm \tau ;X\rangle \langle \pm \tau
;X\!\mid \psi _{\pm }\rangle $ satisfies

\begin{equation}
\label{ec66}\int_{-\infty }^{+\infty }d\tau \,\langle \psi _{\pm }\mid \!\pm
\tau ;X\rangle \langle \pm \tau ;X\!\mid \psi _{\pm }\rangle =\langle \psi
_{\pm }|\psi _{\pm }\rangle =1. 
\end{equation}
Therefore, in a quantum framework, the mean arrival time at $X$ can be
defined consistently by

\begin{equation}
\label{ec62b}\langle t_X\rangle _{\pm }=\pm \langle \psi _{\pm }|\,\hat {%
{\cal T}}(X)\,|\psi _{\pm }\rangle =\int_{-\infty }^{+\infty }d\tau \,\tau
\,\langle \psi _{\pm }\mid \!\pm \tau ;X\rangle \langle \pm \tau ;X\!\mid
\psi _{\pm }\rangle . 
\end{equation}
Accordingly, the probability amplitude $\Psi _{\pm }(t\!=\!\tau ;X)$ of
arriving at $X$ at the instant $t\!=\!\tau $, coming from the left/right,
would be given by

\begin{eqnarray}
\Psi _{\pm }(t\!=\!\tau ;X) & \equiv & \langle \pm \tau ;X|
\psi _{\pm}\rangle =\langle t\!=\!\pm \tau ,\pm ;X|\psi _{\pm }\rangle
\nonumber \\
\label{aplt} &=& h^{-1/2}\!\int_0^\infty \!dp\sqrt{\frac pm}\langle \pm 
p|\psi_{\pm}\rangle \,e^{-i\left(\frac{p^2}{2m}\tau \mp pX\right) /\hbar} 
\end{eqnarray}
and the corresponding probability density takes the form

\begin{equation}
\label{pqua3}|\Psi _{\pm }(t\!=\!\tau ;X)|^2=\!\int_0^\infty \!\!dp^{\prime
}\!\!\int_0^\infty \!\!dp\,\frac{\sqrt{p\,p^{\prime }}}{m\,h}\langle \psi
_{\pm }\mid \!\!\pm p\rangle \langle \pm p^{\prime }|\psi _{\pm }\rangle
\,e^{i\left( \frac{p^2}{2m}-\frac{p^{\prime 2}}{2m}\right) \tau /\hbar
}\,e^{\mp i\left( p-p^{\prime }\right) X/\hbar }. 
\end{equation}

The above formulation can be generalized in order to define a probability
distribution of arrival times at an asymptotic point $X$ behind a
one-dimensional potential barrier. Indeed, provided that the potential $V(x)$
vanishes sufficiently fast, far away from the scattering center, as to
guarantee the validity of the standard scattering formalism, it can be shown
that Eqs. (\ref{ec66})--(\ref{pqua3}) remain formally valid with the only
substitution

\begin{equation}
\label{subs}|\psi _{+}\rangle \rightarrow \frac{|\psi _{{\rm tr}}\rangle }{%
\sqrt{\langle \psi _{{\rm tr}}|\psi _{{\rm tr}}\rangle }}, 
\end{equation}
where the (unnormalized) freely evolving transmitted state $|\psi _{{\rm tr}%
}\rangle $ can be written in terms of the scattering operator ${\hat S}$ as

\begin{equation}
\label{ptri}|\psi _{{\rm tr}}\rangle =\Theta (\hat P)\,{\hat S}\,|\psi _{%
{\rm in}}\rangle =\!\int_0^\infty \!dp\,T(p)\,\langle p|\psi _{{\rm in}%
}\rangle \,|p\rangle . 
\end{equation}
In the above equation $T(p)$ denotes the transmission coefficient
characterizing the potential barrier, and the state vector $|\psi _{{\rm in}%
}\rangle $ [which is assumed to satisfy the identity $|\psi _{{\rm in}%
}\rangle \equiv \Theta (\hat P)\,|\psi _{{\rm in}}\rangle $] represents the
incoming asymptote of the actual scattering state of the particle at $t=0$.
Whenever this latter state $|\psi (0)\rangle $ does not overlap
appreciably with the potential barrier, it becomes physically
indistinguishable from $|\psi _{{\rm in}}\rangle $ and, consequently, it is
not necessary to discriminate between them in practice [\ref{Taylor}].

Since the presence of a potential barrier is not relevant for our purposes
in this work, in what follows we shall restrict ourselves to the free case.
More specifically, we shall consider a freely moving particle characterized
by a state vector $|\psi _{\pm }(t)\rangle $ satisfying Eq. (\ref{psit}).
Nonetheless, this assumption does not imply any loss of generality in
practice since the formulation below can be systematically generalized by
means of the substitution (\ref{subs}).

\vspace{1.2 cm}

\begin{center}
{\large {\bf {III. SEMICLASSICAL LIMIT\vspace{.2 cm}\\}}}
\end{center}

In spite of the fact that the expectation value

\begin{equation}
\label{cachis}\pm \langle \hat J(X)\rangle _{\pm }\equiv \pm \langle \psi
_{\pm }(\tau )|\hat J(X)|\psi _{\pm }(\tau )\rangle =\pm \langle \psi _{\pm
}|\hat J_{{\rm H}}(X,\tau )|\psi _{\pm }\rangle 
\end{equation}
cannot be properly considered as a probability density of arrival times, it
represents, however, a natural quantum version of the corresponding
classical probability density. For this reason it is instructive to 
investigate the connection between such an expectation value and the
quantity $\langle \psi _{\pm }\mid \!\!\pm \tau ;X\rangle \,\langle \pm \tau
;X|\psi _{\pm }\rangle $, which, as stated above, can be interpreted
consistently as a quantum probability density of arrival times. To this end,
by inserting twice the resolution of unity in terms of a momentum basis, we
write

\begin{equation}
\label{esta}\pm \langle \psi _{\pm }|\hat J_{{\rm H}}(X,\tau )|\psi _{\pm
}\rangle =\!\int_0^\infty \!\!dp\!\!\int_0^\infty \!\!dp^{\prime }\,\langle
\psi _{\pm }\mid \!\!\pm p\rangle \langle \pm p^{\prime }|\psi _{\pm
}\rangle \left( \frac{p+p^{\prime }}{2mh}\right) e^{i\left( \frac{p^2}{2m}-%
\frac{p^{\prime 2}}{2m}\right) \tau /\hbar }e^{\mp i(p-p^{\prime })X/\hbar
}. 
\end{equation}
Expressing next the probability amplitude $\langle \pm p|\psi _{\pm }\rangle 
$ in polar form as

\begin{equation}
\label{etcg}\langle \pm p|\psi _{\pm }\rangle =|\langle \pm p|\psi _{\pm
}\rangle |\,e^{i\phi _{\pm }(p)/\hbar } 
\end{equation}
and introducing the functional

\begin{equation}
\label{funt}I_{\pm }\left[ f\right] \equiv \!\int_0^\infty
\!\!dp\,f(p)\,|\langle \pm p|\psi _{\pm }\rangle |\,e^{i\chi _{\pm
}(p)/\hbar }, 
\end{equation}
where the phase $\chi _{\pm }(p)$ is defined as

\begin{equation}
\label{mfal}\chi _{\pm }(p)\equiv \phi _{\pm }(p)-\frac{p^2}{2m}\tau \pm pX, 
\end{equation}
we can finally rewrite Eq. (\ref{esta}) in the form

\begin{equation}
\label{jhcc}\pm \langle \psi _{\pm }|\hat J_{{\rm H}}(X,\tau )|\psi _{\pm
}\rangle =\frac 1{mh}\frac 12\left( I_{\pm }^{*}\left[ p\right] I_{\pm
}\left[ 1\right] +{\rm c.c.}\right) . 
\end{equation}
On the other hand, the probability density (\ref{pqua3}) (which, unlike the
probability current, is manifestly positive definite) can be written in a
completely analogous manner as

\begin{equation}
\label{pmi}\langle \psi _{\pm }\mid \!\!\pm \tau ;X\rangle \,\langle \pm
\tau ;X|\psi _{\pm }\rangle =\frac 1{mh}\left( I_{\pm }^{*}\left[ \sqrt{p}%
\right] I_{\pm }\left[ \sqrt{p}\right] \right) . 
\end{equation}

The two expressions (\ref{jhcc}) and (\ref{pmi}) are especially
suitable for investigating the semiclassical limit $\hbar \rightarrow 0$.
Indeed, in this limit the asymptotic expansion of the Fourier-type integral 
(\ref{funt}) is given, to leading order, by [\ref{asymp}]

\begin{equation}
\label{astt}I_{\pm }\left[ f\right] \sim \,e^{i\frac \pi 4\,{\rm sgn}[\chi
_{\pm }^{\prime \prime }(p_0)]}\,e^{i\chi _{\pm }(p_0)/\hbar
}f(p_0)\,|\langle \pm p_0|\psi _{\pm }\rangle |\,\sqrt{\frac h{|\chi _{\pm
}^{\prime \prime }(p_0)|}}, 
\end{equation}
where $p_0$ is the stationary point of the phase $\chi _{\pm }(p)$, defined
implicitly by

\begin{equation}
\label{ftae}\chi _{\pm }^{\prime }(p_0)\equiv \phi _{\pm }^{\prime }(p_0)-%
\frac{p_0}m\tau \pm X=0, 
\end{equation}
and primes are used to denote differentiation with respect to the momentum $%
p $.

By substituting Eq. (\ref{astt}) into Eqs. (\ref{jhcc}) and (\ref{pmi}) we
obtain, to leading order as $\hbar \rightarrow 0$,

\begin{equation}
\label{cont1}\langle \psi _{\pm }\mid \!\!\pm \tau ;X\rangle \,\langle \pm
\tau ;X|\psi _{\pm }\rangle \sim \pm \langle \psi _{\pm }|\hat J_{{\rm H}%
}(X,\tau )|\psi _{\pm }\rangle \sim \frac{p_0}m|\psi _{\pm }(X,\tau )|^2. 
\end{equation}
In the last step of the above formula we have used that

\begin{equation}
\label{ponla}\psi _{\pm }(X,\tau )=\langle X|e^{i\frac{\!\hat P^2}{2m}\tau
/\hbar }|\psi _{\pm }\rangle =h^{-1/2}\,\!I_{\pm }\left[ 1\right] , 
\end{equation}
so that, as $\hbar \rightarrow 0$ we have

\begin{equation}
\label{afn}|\psi _{\pm }(X,\tau )|^2\sim \frac{|\langle \pm p_0|\psi _{\pm
}\rangle |^2}{|\chi _{\pm }^{^{\prime \prime }}(p_0)|}. 
\end{equation}

The information contained in Eq. (\ref{cont1}) represents the main result of
this section. This equation reflects that the proposed probability density
of arrival times $|\langle \pm \tau ;X|\psi _{\pm }\rangle |^2$ coincides,
in the semiclassical limit, with the quantum probability current $\pm
\langle \psi _{\pm }|\hat J_{{\rm H}}(X,\tau )|\psi _{\pm }\rangle $, which
in turn is given in this limit by the product of the probability density $%
|\psi _{\pm }(X,\tau )|^2$ and the group velocity $p_0/m$ (and,
consequently, becomes a positive quantity). This fact suggests that the
probability density defined in Eq. (\ref{pqua3}) is nothing but a quantum
version of the modulus of the classical average current $\langle J(X)\rangle 
$, which, as already stated, plays the role of a probability distribution of
arrival times at $X$ for a classical statistical ensemble of particles
moving along a well-defined spatial direction. In the next section we shall
see that this is indeed the case.

\vspace{1.2 cm}

\begin{center}
{\large {\bf {IV. POSITIVE-DEFINITE CURRENT\vspace{.2 cm}\\}}}
\end{center}

Let us concentrate on Eq. (\ref{ec69}), which defines the eigenstates of the
operator $\hat {{\cal T}}(X)$ in terms of a basis of momentum eigenstates.
By introducing the self-adjoint operator

\begin{equation}
\label{hbm}\sqrt{|\hat P\,|}\equiv \int_{-\infty }^{+\infty }dp\sqrt{|p\,|}%
\,|p\rangle \langle p|, 
\end{equation}
we can express $|\tau ;X\rangle $ in an alternative form that exhibits no
explicit dependence on any particular representation and proves to be most
convenient for our purposes. Indeed,

\begin{equation}
\label{indp}|\tau ;X\rangle =\!\int_{-\infty }^{+\infty }\!dp\,\langle p|\,%
\sqrt{\frac{|\hat P\,|}m}\,e^{i\,{\rm sgn}(\hat P\,)\frac{\hat P\,^2}{2m}%
\tau /\hbar }\,|X\rangle \,|p\rangle =\sqrt{\frac{|\hat P\,|}m}\,e^{i\,{\rm %
sgn}(\hat P\,)\frac{\hat P\,^2}{2m}\tau /\hbar }\,|X\rangle . 
\end{equation}
Correspondingly, the probability amplitude for particle detection at the
spatial point $X$ [coming from the left($+$)/right($-$)] at time $t=\tau $
takes the form [Eq. (\ref{aplt})]

\begin{equation}
\label{ampl2}\Psi _{\pm }(t\!=\!\tau ;X)=\langle X|\,\sqrt{\frac{|\hat P\,|}m%
}\,e^{-i\,\frac{\hat P\,^2}{2m}\tau /\hbar }\,|\psi _{\pm }\rangle =\langle
X|\,\sqrt{\frac{|\hat P\,|}m}\,|\psi _{\pm }(\tau )\rangle . 
\end{equation}
This equation shows that the probability amplitude of arriving at $X$ at
time $\tau $ is nothing but the probability amplitude of finding the state $%
\sqrt{|\hat P\,|/m}\,|X\rangle $ in the (Schr\"odinger) state vector $|\psi
_{\pm }(\tau )\rangle $ characterizing the particle dynamics at $t=\tau $.
On the other hand, the corresponding probability density reads

\begin{equation}
\label{pqua}\left. P_X^{(\pm )}(\tau )\right| _{{\rm quant}}=|\Psi _{\pm
}(t\!=\!\tau ;X)|^2=\langle \psi _{\pm }(\tau )|\,\sqrt{\frac{|\hat P\,|}m}%
\,\delta (\hat X\,-X)\,\sqrt{\frac{|\hat P\,|}m}\,|\psi _{\pm }(\tau
)\rangle , 
\end{equation}
where use has been made of the identity $|X\rangle \langle X|\equiv \delta
(\hat X\,-X)$.

Before proceeding further it is convenient to consider the {\em normalized}
probability distribution of arrival times at the point $X$ for a classical
statistical ensemble of {\em free} particles of mass $m$, coming either from
the left $(p>0)$ or from the right $(p<0)$. Such a probability distribution
can be obtained from Eq. (\ref{jota}). Indeed, by defining $J^{(+)}(X)$ as
the modulus of the classical current

\begin{equation}
\label{xwx7}J^{(+)}(X)\equiv |J(X)|= |p|/m\,\delta (x-X)  
\end{equation}
and using a convenient notation, it takes the form

\begin{equation}
\label{jota2}\left. P_X^{(\pm )}(\tau )\right| _{{\rm class}}=\pm \langle
J(X)\rangle _{\pm }=\!\int \!\!\int \!f_{\pm }(x,p,\tau )\,\frac{|p|}%
m\,\delta (x-X)\,dx\,dp = \langle J^{(+)}(X)\rangle _{\pm }, 
\end{equation}
where the phase space distribution function satisfies the identity

\begin{equation}
\label{fpm1}\!f_{\pm }(x,p,t)\equiv \Theta (\pm p)\,\!f_{\pm }(x,p,t) 
\end{equation}
and the modulus in the integrand of Eq. (\ref{jota2}) comes from the
normalization factor (which, in the free case, takes the value $\pm 1$).

A comparison between expressions (\ref{pqua}) and (\ref{jota2}) shows that
the quantum probability density of arrival times defined above [Eq. (\ref
{pqua})] can be considered as a quantum version of the corresponding
classical expression, obtained by associating to the average of the
classical positive current $J^{(+)}(X)\equiv |J(X)|$ the expectation value
of the positive definite current operator $\hat J^{(+)}(X)$

\begin{equation}
\label{corr1}J^{(+)}(X)\equiv \frac{|p|}m\,\delta (x-X)\rightarrow \hat
J^{(+)}(X)\equiv \sqrt{\frac{|\hat P\,|}m}\,\delta (\hat X\,-X)\,\sqrt{\frac{%
|\hat P\,|}m}. 
\end{equation}
It should be noted, however, that the relation existing between the
classical current $J(X)$ and the corresponding quantum operator $\hat J(X)$
[given by Eq. (\ref{curop})] is somehow different from that existing between 
$J^{(+)}(X)$ and $\hat J^{(+)}(X)$. Indeed, $\hat J(X)$ can also be
considered as the quantum operator corresponding to the classical current $%
J(X)$ by virtue of the Weyl-Wigner quantization rule, whereas the same does
not hold true for the positive current defined above.

The Weyl-Wigner quantization rule is a mapping that associates with every
phase-space function $g(x,p)$ a quantum operator $\hat G(\hat X,\hat P)$
with an expectation value satisfying

\begin{equation}
\label{unms1}\langle \hat G(\hat X,\hat P)\rangle =\!\int \!\!\int \!f_{{\rm %
W}}(x,p)\,g(x,p)\,dx\,dp, 
\end{equation}
where the {\em Wigner function} $f_{{\rm W}}(x,p)$ plays the role of a
quasiprobability distribution function in phase space [\ref{Wigner2}] and
can be expressed in terms of the quantum density operator $\hat \rho $
characterizing the physical system as [\ref{Cohen}]

\begin{equation}
\label{otms}f_{{\rm W}}(x,p)=\frac 1{4\pi ^2}\!\int \!\!\int \!\!\int
\!\left\langle q+\frac{\tau \hbar }2\left| \hat \rho \right| q-\frac{\tau
\hbar }2\right\rangle e^{-i[\theta (x-q)+\tau p]}\,d\theta \,d\tau \,dq. 
\end{equation}
By substituting Eq. (\ref{otms}) into Eq. (\ref{unms1}) it can be shown that 
$\hat G(\hat X,\hat P)$ is given by

\begin{equation}
\label{epg1}\hat G(\hat X,\hat P)=\frac 1{4\pi ^2}\!\int \!\!\int \!\!\int
\!\!\int \!g(x,p)e^{i[\theta (\hat X-q)+\tau (\hat P-p)]}\,dx\,dp\,d\theta
\,d\tau , 
\end{equation}
and taking $g(x,p)\equiv p/m\,\delta (x-X)$ in the integrand of Eq. (\ref
{epg1}) one arrives, after some algebra, at the current operator $\hat J(X)$
defined by Eq. (\ref{curop}). Even though a similar relation does not exist
for $J^{(+)}(X)\equiv |J(X)|$ it still holds true that the positive-definite
operator $\hat J^{(+)}(X)$ represents a natural quantum version of the
modulus of the classical current. Accordingly, for free particles
propagating along a well-defined spatial direction the probability density
of the time of arrival at a given point $X$ at time $t=\tau $ can be
defined consistently, within both a classical and a quantum-mechanical
framework, as the instantaneous mean value of the modulus of the current

\begin{equation}
\label{pcl1}\left. P_X^{(\pm )}(\tau )\right| _{{\rm class}}=\langle
J^{(+)}(X)\rangle _{\pm } ,
\end{equation}

\begin{equation}
\label{pq1}\left. P_X^{(\pm )}(\tau )\right| _{{\rm quant}}=\langle \psi
_{\pm }\mid \!\!\pm \tau ;X\rangle \,\langle \pm \tau ;X|\psi _{\pm }\rangle
=\langle \psi _{\pm }(\tau )|\hat J^{(+)}(X)|\psi _{\pm }(\tau )\rangle . 
\end{equation}

We shall next concentrate on the operator $\hat {{\cal T}}(X)$. From the
definition (\ref{ec68}) one finds, taking Eq. (\ref{indp}) into account, 
that $\hat {{\cal T}}(X)$ satisfies

\begin{equation}
\label{ecas}\pm \,\Theta (\pm \hat P)\,\hat {{\cal T}}(X)\,\Theta (\pm \hat
P)=\,\!\Theta (\pm \hat P)\left[ \int_{-\infty }^{+\infty }\!d\tau \,\tau
\,\hat J_{{\rm H}}^{(+)}(X,\tau )\right] \Theta (\pm \hat P), 
\end{equation}
where $\hat J_{{\rm H}}^{(+)}(X,\tau )$ denotes the positive current in the
Heisenberg picture

\begin{equation}
\label{jph}\hat J_{{\rm H}}^{(+)}(X,\tau )=e^{i\hat H_0\tau /\hbar }\,\hat
J^{(+)}(X)\,e^{-i\hat H_0\tau /\hbar }. 
\end{equation}
Accordingly, the mean arrival time at $X \,$ [ Eq. (\ref{ec62b})] can be
expressed as

\begin{equation}
\label{ttem}\langle t_X\rangle _{\pm }=\pm \langle \psi _{\pm }|\,\hat {%
{\cal T}}(X)\,|\psi _{\pm }\rangle =\!\int_{-\infty }^{+\infty }\!d\tau
\,\tau \,\langle \psi _{\pm }(\tau )|\hat J^{(+)}(X)|\psi _{\pm }(\tau
)\rangle . 
\end{equation}
The above equation gives the mean arrival time in a form that can be
recognized as a quantum version of its classical counterpart in terms of the
probability distribution (\ref{jota2}). Indeed, the positive current $%
\langle \psi _{\pm }(\tau )|\hat J^{(+)}(X)|\psi _{\pm }(\tau )\rangle $
enters Eq. (\ref{ttem}) as a probability density of the time of arrival at $%
X $. It should be stressed that contrary to what happens with Eq. (\ref{ec62}%
), by virtue of Eq. (\ref{ecas}) the above equation is valid for {\em any} $%
|\psi _{\pm }(\tau )\rangle $ satisfying the condition (\ref{psit}). When in
addition to this latter condition it is also satisfied that

\begin{equation}
\label{ec51}{\rm \lim _{p\rightarrow \pm \infty }\;}\langle p|\psi _{\pm
}\rangle =0,
\;\;\;\;\;\;\;\;\;\;
{\rm \lim _{p\rightarrow 0}\;\,}p^{-1}\,\langle p|\psi _{\pm }\rangle =0, 
\end{equation}
it can be shown, after some algebra, that the mean arrival time $%
\langle t_X\rangle _{\pm }$ can also be expressed in the 
alternative forms

\begin{equation}
\label{jotr}\langle t_X\rangle _{\pm }=\pm \!\int_{-\infty }^{+\infty
}\!d\tau \,\tau \,\langle \psi _{\pm }(\tau )|\hat J(X)|\psi _{\pm }(\tau
)\rangle , 
\end{equation}

\begin{equation}
\label{taha}\langle t_X\rangle _{\pm }=\langle \psi _{\pm }|\,\frac 12\left(
\left[ X-\hat X_{{\rm H}}(0)\right] \frac m{\hat P_{{\rm H}}(0)\,}+\frac
m{\hat P_{{\rm H}}(0)\,}\,\left[ X-\hat X_{{\rm H}}(0)\right] \right) |\psi
_{\pm }\rangle , 
\end{equation}

\begin{equation}
\label{trov}\langle t_X\rangle _{\pm }=\langle \psi _{\pm }|\,\sqrt{\frac
m{\hat P_{{\rm H}}(0)}}\left[ X-\hat X_{{\rm H}}(0)\right] \sqrt{\frac
m{\hat P_{{\rm H}}(0)}}\,|\psi _{\pm }\rangle . 
\end{equation}
In the derivation of the above formulas use has been made of the fact that
the position operator transforms under spatial translations as

\begin{equation}
\label{venga}e^{-i\hat PX/\hbar }\,\hat X\,\,e^{+i\hat PX/\hbar }=(\hat
X\,-X) 
\end{equation}
and that, at $t=0$, quantum operators in the Schr\"odinger picture become
indistinguishable from those in the Heisenberg picture, so that, in
particular, we have $\hat X=\hat X_{{\rm H}}(0)$ and $\hat P=\hat P_{{\rm H}%
}(0)$.

Equations (\ref{taha}) and (\ref{trov}) give the mean arrival time at $X$ in
terms of the operators $\hat T(X)$ introduced by Grot {\em et al.} and by
Aharonov and Bohm [Eqs. (\ref{trove1}) and (\ref{tahar1}), respectively]. A
connection between these operators and the self-adjoint ''time'' operator $%
\hat {{\cal T}}(X)$ can be derived from a comparison with Eq. (\ref{ttem}%
). Indeed, as long as conditions (\ref{psit}) and (\ref{ec51}) are
satisfied, the expectation value of $\hat {{\cal T}}(X)$ coincides with
the expectation values of the operators $\hat T(X)$, which have the interest
that they can be obtained by quantizing the classical expression $%
t(X)=m[X-x(0)]/p(0)$ according to different standard quantization (ordering)
rules.

Under the same conditions, the expression of $\langle t_X\rangle _{\pm }$
given by Eq. (\ref{ttem}) becomes also indistinguishable from that given by
Eq. (\ref{jotr}), which involves the usual quantum probability current and
has been frequently used, in practice, as a quantum definition for the mean
arrival time. This fact may provide additional justification for the latter
expression, whose validity in a quantum framework might, in principle, be
questionable. Indeed, despite the formal analogy between Eqs. (\ref{ttem})
and (\ref{jotr}), they have a somewhat different physical meaning: While the
positive current $\hat J^{(+)}(X)$ is a positive-definite operator and its
expectation value enters Eq. (\ref{ttem}) playing the role of a probability
density, the expectation value of the usual probability current $\hat J(X)$
can take negative values and consequently cannot be interpreted as a
probability distribution of arrival times. Since usually this has been the
case, however, it is instructive analyzing under what circumstances the
expectation values of $\hat J(X)$ and $\hat J^{(+)}(X)$ become
indistinguishable. This will be the aim of the next section.

\vspace{1.2 cm}

\begin{center}
{\large {\bf {V. PROBABILITY CURRENT VERSUS POSITIVE CURRENT \vspace{.2 cm}\\%
}}}
\end{center}

In the preceding section we have seen that for state vectors satisfying 
Eqs. (\ref{psit}) and (\ref{ec51}) the mean arrival time 
$\langle t_X\rangle _{\pm }$ can be equally calculated by using the positive 
current or the probability current [Eqs. (\ref{ttem}) and (\ref{jotr}),  
respectively]. Furthermore, by using

\begin{equation}
\label{delt2}\int_{-\infty }^{+\infty }\!\!d\tau \,e^{i\left( \frac{p^2}{2m}-%
\frac{p^{\prime 2}}{2m}\right) \tau /\hbar }=\frac m{|p|}\,\delta (p^{\prime
}-p)+\frac m{|p|}\,\delta (p^{\prime }+p) 
\end{equation}
in Eqs. (\ref{pqua3}) and (\ref{esta}), we obtain

\begin{equation}
\label{mfte}\int_{-\infty }^{+\infty }\!d\tau \,\langle \psi _{\pm }(\tau
)|\hat J^{(+)}(X)|\psi _{\pm }(\tau )\rangle =\pm \!\int_{-\infty }^{+\infty
}\!d\tau \,\langle \psi _{\pm }(\tau )|\hat J(X)|\psi _{\pm }(\tau )\rangle
, 
\end{equation}
so that the total probability of arriving at $X$ (at any instant of time)
can also be calculated in terms of $\hat J^{(+)}(X)$ or $\hat J(X)$. Despite
the interchangeable role that these quantities play in the above integral
expressions, when the state vector describing the quantum particle exhibits
a considerable spread in momentum space the contribution of quantum
interference effects may be important and the expectation values of $\hat
J(X)$ and $\hat J^{(+)}(X)$ can be appreciably different. This is most
easily seen by considering a superposition of two nonoverlapping wave 
packets with well-defined momentum. Specifically, we shall consider a state 
vector $|\psi _{+}\rangle $ given by

\begin{equation}
\label{supr}|\psi _{+}\rangle =\alpha _1\,|\psi _1\rangle +\alpha _2\,|\psi
_2\rangle , 
\end{equation}
where the coefficients $\alpha _1,\alpha _2\,$ are real and $\langle
p|\psi _j\rangle $ $(j=1,2)$ are assumed to be minimum Gaussian wave packets
centered at the spatial point $x_0$, with momentum spread $\Delta p$ and
average momentum $p_j$, respectively,

\begin{equation}
\label{gausp}\langle p|\psi _j\rangle =\left[ 2\pi (\Delta p)^2\right]
^{-1/4}\exp \left[ -\left( \frac{p-p_j}{2\Delta p}\right) ^2-i\frac{px_0}%
\hbar \right] . 
\end{equation}
We take $p_2>p_1>0$ and $\Delta p\ll (p_2-p_1)$ in order to guarantee that
the above two wave packets do not overlap appreciably. Under these
assumptions, the probability current $\langle \psi _{+}|\hat J_{{\rm H}%
}(X,\tau )|\psi _{+}\rangle $ and the probability density $\langle \psi
_{+}|\hat J_{{\rm H}}^{(+)}(X,\tau )|\psi _{+}\rangle $ can be written,
respectively, as

\begin{equation}
\label{prim}\langle \psi _{+}|\hat J_{{\rm H}}(X,\tau )|\psi _{+}\rangle
=\frac 1{mh}\frac 12\left( I^{*}\left[ p\right] I\left[ 1\right]
+{\rm c.c.}\right) , 
\end{equation}

\begin{equation}
\label{segu}\langle \psi _{+}|\hat J_{{\rm H}}^{(+)}(X,\tau )|\psi
_{+}\rangle =\frac 1{mh}\left( I^{*}\left[ \sqrt{p}\right] I\left[ \sqrt{p}%
\right] \right) , 
\end{equation}
where now the functional $I\left[ f\right] $ is given by $I\left[ f\right]
\equiv I_1[f]+I_2[f]$ with

\begin{equation}
\label{ijot}I_j[f]\equiv \alpha _j\,\left[ 2\pi (\Delta p)^2\right]
^{-1/4}\!\int_0^\infty \!\!dp\,f(p)\,\exp \left[ -\left( \frac{p-p_j}{%
2\Delta p}\right) ^2-i\frac{px_0}\hbar \right] \exp \left[ -\frac{p^2}{2m}%
\frac \tau \hbar +i\frac{pX}\hbar \right] . 
\end{equation}

To obtain an analytical estimation (as a function of $\tau $) for the
probability current and for the probability density of arrival times [as
given by Eqs. (\ref{prim}) and (\ref{segu}), respectively] we shall next
consider the asymptotic expansion of the above integral. To leading order as 
$\Delta p\rightarrow 0$, we have [\ref{asymp}]

\begin{equation}
\label{asy4}I_j[f]\sim \alpha _j\,\sqrt{2}\left[ 2\pi (\Delta p)^2\right]
^{1/4}f(p_j)\exp \left[ -\frac{p_j^2}{2m}\frac \tau \hbar
-i\,p_j(x_0-X)/\hbar \right] +O\left[ \frac{(\Delta p)^3}{\sqrt{\Delta p}}%
\right] . 
\end{equation}
By substituting Eq. (\ref{asy4}) into Eqs. (\ref{prim}) and (\ref{segu}),
one obtains, after some algebra, the asymptotic expressions

\begin{eqnarray}
\langle \psi _{+}|\hat J_{{\rm H}}(X,\tau )|\psi _{+}\rangle \sim 
\frac{2\sqrt{2\pi }}{mh}\Delta p \left( \alpha _1^2p_1+\alpha _2^2p_2+
\alpha _1\alpha_2(p_1+p_2) \right. \nonumber \\
\label{cosj} \times \left. \cos \left[ (p_2^2-p_1^2)\tau /2m+(p_2-p_1)
(x_0-X)\right] /\hbar+O\left[ (\Delta p)^2\right] \right), 
\end{eqnarray}

\begin{eqnarray}
\langle \psi _{+}|\hat J_{{\rm H}}^{(+)}(X,\tau )|\psi _{+}\rangle 
\sim  \frac{2\sqrt{2\pi }}{mh}\Delta p \left( \alpha _1^2p_1+
\alpha _2^2p_2+\alpha_1\alpha _2\sqrt{p_1p_2} \right. \nonumber \\
\label{cosjp} \times \left. \cos \left[ (p_2^2-p_1^2)\tau /2m+(p_2-p_1)
(x_0-X)\right]/\hbar +O\left[ (\Delta p)^2\right] \right). 
\end{eqnarray}

From Eq. (\ref{cosj}) we see that the probability current can take negative
values whenever the interference term dominates over both the first and the
second one. It is not hard to see that this is the case when it holds that

\begin{equation}
\label{cond1}1\ll \frac{\alpha _1}{\alpha _2}\ll \frac{p_2}{p_1}, 
\end{equation}
and, under these circumstances, Eq. (\ref{cosj}) is given, to a good
approximation, by

\begin{equation}
\label{fteta}\langle \psi _{+}|\hat J_{{\rm H}}(X,\tau )|\psi _{+}\rangle
\sim \frac{2\sqrt{2\pi }}{mh}\Delta p\,\alpha _1\alpha _2\,p_2\cos \left[
p_2^2\tau /2m+p_2(x_0-X)\right] /\hbar +O\left[ (\Delta p)^3\right] . 
\end{equation}
In contrast, the probability density $\langle \psi _{+}|\hat J_{{\rm H}%
}^{(+)}(X,\tau )|\psi _{+}\rangle $ remains always positive. This fact,
which is evident from Eq. (\ref{segu}), can also be verified from the
asymptotic expression (\ref{cosjp}) by noting that the dominance of the
interference term would require that

\begin{equation}
\label{unms2}1\ll \frac{\alpha _1}{\alpha _2}\sqrt{\frac{p_1}{p_2}}\ll 1, 
\end{equation}
which is obviously impossible.

On the other hand, it can be readily verified that for nonnormalizable
states with a well-defined momentum $|\psi _{\pm }\rangle =\,|p\rangle $ ($%
p\neq 0$), the quantity $\langle \psi _{\pm }(\tau )|\hat J^{(+)}(X)|\psi
_{\pm }(\tau )\rangle $ becomes indistinguishable from the usual probability
current $\pm \langle \psi _{\pm }(\tau )|\hat J(X)|\psi _{\pm }(\tau
)\rangle $. One expects this fact to be also true for normalizable states
describing particles with a highly defined momentum, and this is indeed the
case as can be inferred from the asymptotic behavior (as the momentum
uncertainty approaches zero) of the integrals defining the expectation
values of $\hat J(X)$ and $\hat J^{(+)}(X)$ [see Eqs. (\ref{cosj}) and (\ref
{cosjp}) above and take $\alpha _2\equiv 0$].

In scattering problems one is usually concerned with particles propagating
with a well-defined velocity toward a localized interaction center. Such
particles are characterized by quantum states highly concentrated in
momentum space about a certain momentum $p_0\neq 0$. Under these conditions,
the expectation values of $\hat J(X)$ and $\hat J^{(+)}(X)$ coincide to a
good approximation, so that the probability current yields essentially
correct results for the probability density of the time of arrival at a
given point. This fact provides a justification for the use of $\hat J(X)$
in this kind of problems.

Finally, it is worth noting that Eq. (\ref{jotr}), which gives the mean
arrival time $\langle t_X\rangle _{\pm }$ in terms of the probability
current, is applicable even for state vectors having a large momentum
uncertainty. Indeed, its validity only requires the fulfillment of
conditions (\ref{psit}) and (\ref{ec51}).

\vspace{1.2 cm}

\begin{center}
{\large {\bf {VI. CONCLUSION \vspace{.2 cm}\\}}}
\end{center}

Quantum theories are assumed to be more fundamental in nature than the
corresponding classical theories. Consequently, it is possible, in
principle, to define quantum quantities without resorting to the 
correspondence principle. In practice, however, the correspondence principle 
proves to be extremely useful in the construction of the quantum counterpart 
of a certain classical quantity. For instance, the canonical quantization 
method represents an invaluable tool for the construction of quantum field
theories. Furthermore, one usually gets a better understanding of a quantum
theory when a clear and unambiguous relationship can be established between
quantum and classical quantities.

In this paper we have been particularly interested in investigating the
connection between the expressions previously proposed in Ref. [\ref{VDB4}]
for the probability distribution of the time of arrival at a given spatial
point and their corresponding classical counterparts. In particular, we have
shown that, in the semiclassical limit $\hbar \rightarrow 0$, the proposed
probability density of arrival times coincides with the modulus of the 
quantum probability current. Indeed, Eq. (\ref{cont1}) can be rewritten in 
the form

\begin{equation}
\label{cont12}|\langle \pm \tau ;X|\psi _{\pm }\rangle |^2\sim |\langle \psi
_{\pm }(\tau )|\hat J(X)|\psi _{\pm }(\tau )\rangle |. 
\end{equation}
This result has the interest that, at a classical level, the current of a
statistical ensemble of particles propagating along a well-defined spatial
direction plays the role of a probability distribution of arrival times.
Therefore, Eq. (\ref{cont12}) reflects that the quantity $|\langle \pm \tau
;X|\psi _{\pm }\rangle |^2$ has the correct semiclassical limit and
suggests that it represents a quantum version of the corresponding classical
expression. We have explicitly shown that this is the case by expressing the
probability distribution $|\langle \pm \tau;X|\psi _{\pm }\rangle |^2$ as
the expectation value of a certain positive definite current operator.
Indeed, by making use of the fact that the probability amplitude of arriving
at $X$ at time $\tau $ coincides with the probability amplitude of finding
the state vector $\sqrt{|\hat P\,|/m}\,|X\rangle $ in the Schr\"odinger
state of the particle $|\psi _{\pm }(\tau )\rangle $, one can write the
corresponding probability density in the form

\begin{equation}
\label{pq12}\left. P_X^{(\pm )}(\tau )\right| _{{\rm quant}}\equiv |\langle
\pm \tau ;X|\psi _{\pm }\rangle |^2=\langle \psi _{\pm }(\tau )|\hat
J^{(+)}(X)|\psi _{\pm }(\tau )\rangle , 
\end{equation}
where the positive-definite operator

\begin{equation}
\label{pqua7}\hat J^{(+)}(X)\,=\,\sqrt{\frac{|\hat P\,|}m}\,\delta (\hat
X\,-X)\,\sqrt{\frac{|\hat P\,|}m}\,, 
\end{equation}
can be immediately recognized as a straightforward quantum version of the
modulus of the classical current $|J(X)|=|p|/m\,\delta (x-X)$. The existence 
of a remarkable formal analogy between the corresponding classical and 
quantum expressions is therefore apparent: For particles propagating
along a well-defined spatial direction, the probability distribution of the 
time of arrival at a given point can be inferred, within both a classical 
and a quantum framework, from the mean value of the modulus of the current.

On the other hand, for normalizable states satisfying the identity

\begin{equation}
\label{chac}|\psi _{\pm }(t)\rangle \equiv \Theta (\pm \hat P)\,|\psi _{\pm
}(t)\rangle  
\end{equation}
and vanishing faster than $p$ as $p$ approaches zero, the mean arrival time
at $X$, $\langle t_X\rangle _{\pm }$, can be equally calculated in terms of
the positive-definite current $\hat J^{(+)}(X)$ or in terms of the standard
probability current. This interchangeable role is not restricted to the mean
arrival time. Indeed, we have seen that for physical states with a
sufficiently well-defined momentum the expectation values of $\hat J(X)$ and 
$\hat J^{(+)}(X)$ become indistinguishable, so that the probability current
yields essentially correct results for the probability density of the time
of arrival at a given point. This fact may provide a justification for the
common practice of using the expectation value of $\hat J(X)$ in this kind
of problem.

Furthermore, under the same above conditions, the expectation value of
the self-adjoint ''time'' operator $\hat {{\cal T}}(X)$ coincides with the
expectation values of the operators $\hat T(X)$ introduced by Grot {\em et
al.} and by Aharonov and Bohm [Eqs. (\ref{trove1}) and (\ref{tahar1}),
respectively], which have the interest that they can be considered as
straightforward quantum versions of the classical expression $%
t(X)=m[X-x(0)]/p(0)$.

In summary, we have shown that the formalism developed in Ref. [\ref{VDB4}]
for the time of arrival of a quantum particle at a given spatial point can
be reformulated in a form that exhibits a remarkable formal analogy with
the corresponding classical formulation.

\vspace{1.2cm}

{\bf Note added in proof:}

\vspace{.2 cm}

As stated in the Introduction, the idea that according to standard quantum
mechanics measuring results of physical quantities can be inferred from the
spectral decomposition of a certain self-adjoint operator, without having to
make reference to the specific properties of the measuring device involved,
plays a central role in our treatment. In this regard it should be 
mentioned that a completely different view is developed in a recent preprint 
by Aharonov {\em et al. } [\ref{NEW1}]. These authors, by explicitly 
modelling the measuring device, arrive at the conclusion that the time 
of arrival cannot be precisely defined and measured in quantum mechanics. 
In the view of the present author, however, this pessimistic conclusion 
can be mitigated, in part, by the assumptions on which it is based 
[\ref{NEW2}].

On the other hand, in connection with Eq. (\ref{cont12}) [which is only 
valid in the semiclassical limit] it is interesting to note that, as shown 
by McKinnon and Leavens [\ref{Leav3}], within the framework of Bohmian 
mechanics the modulus of the quantum probability current provides a 
definition for the probability density of arrival times which is of general 
applicability [\ref{NEW3}]. Therefore, while according to Eq. (\ref{pq12}) 
within conventional quantum mechanics the probability density of the time 
of arrival can be defined as the mean value of the modulus of the current, 
within Bohmian mechanics it is the modulus of the mean value of the 
current the relevant quantity (the range of applicability of the two 
expressions is different, however).

\vspace{1.2 cm}

\begin{center}
{\large {\bf {ACKNOWLEDGMENT \vspace{.2 cm}\\}}}
\end{center}

This work has been supported by Gobierno Aut\'onomo de Canarias (Project No.
PI 2/95).

\newpage
\vspace{1.4 cm}

\begin{center}
{\large {\bf REFERENCES \vspace{.6cm}}}
\end{center}

\begin{enumerate}
\item  \label{Du} R. S. Dumont and T. L. Marchioro II, Phys. Rev. A {\bf 47}%
, 85 (1993).

\item  \label{Leav2} C. R. Leavens, Phys. Lett. A {\bf 178}, 27 (1993).

\item  \label{Leav3} W. R. McKinnon and C. R. Leavens, Phys. Rev. A {\bf 51}%
, 2748 (1995).

\item  \label{Mug} J. G. Muga, S. Brouard, and D. Mac\'\i as, Ann. Phys.
(N.Y.) {\bf 240}, 351 (1995).

\item  \label{Pauli} W. Pauli, in {\it Encyclopaedia of Physics}, edited by
S. Flugge (Springer, Berlin, 1958), Vol. 5/1, p. 60.

\item  \label{Allco} G. R. Allcock, Ann. Phys. (N.Y.) {\bf 53}, 253 (1969); 
{\bf 53}, 286 (1969); {\bf 53}, 311 (1969).

\item  \label{VDB4} V. Delgado and J. G. Muga, Phys. Rev. A {\bf 56}, 3425 
(1997) (quant-ph/9704010).

\item  \label{Kij} J. Kijowski, Rep. Math. Phys. {\bf 6}, 361 (1974).

\item  \label{Grot} N. Grot, C. Rovelli, and R. S. Tate, Phys. Rev. A {\bf 54%
}, 4676 (1996).

\item  \label{Mielnik} B. Mielnik, Found. Phys. {\bf 24}, 1113 (1994).

\item  \label{Busch} P. Busch, M. Grabowski, and P. J. Lahti, Phys. Lett. A 
{\bf 191}, 357 (1994).

\item  \label{Gian} R. Giannitrapani, Int. J. Theor. Phys. {\bf 36}, 1601
(1997).

\item  \label{Lan} M. B\"uttiker and R. Landauer, Phys. Rev. Lett. {\bf 49},
1739 (1982).

\item  \label{Rev} For recent reviews on the subject see (a) E. H. Hauge and
J. A. St\/ovneng, Rev. Mod. Phys. {\bf 61}, 917 (1989); (b) M. B\"uttiker,
in {\it Electronic Properties of Multilayers and Low-Dimensional
Semiconductor Structures}, edited by J. M. Chamberlain {\it et al.} (Plenum,
New York, 1990), p. 297; (c) R. Landauer, Ber. Bunsenges. Phys. Chem. {\bf 95%
}, 404 (1991); (d) C. R. Leavens and G. C. Aers, in {\it Scanning Tunneling
Microscopy III}, edited by R. Wiesendanger and H. J. G\"utherodt (Springer,
Berlin, 1993), pp. 105--140; (e) R. Landauer and T. Martin, Rev. Mod. Phys. 
{\bf 66}, 217 (1994).

\item  \label{Bu} M. B\"uttiker, Phys. Rev. B {\bf 27}, 6178 (1983).

\item  \label{San} S. Brouard, R. Sala, and J. G. Muga, Phys. Rev. A {\bf 49}%
, 4312 (1994).

\item  \label{Reca} V. S. Olkhovsky and E. Recami, Phys. Rep. {\bf 214}, 339
(1992).

\item  \label{Leav} C. R. Leavens, Solid State Commun. {\bf 85}, 115 (1993); 
{\bf 89}, 37 (1993).

\item  \label{VDB1} V. Delgado, S. Brouard, and J. G. Muga, Solid State
Commun. {\bf 94}, 979 (1995).

\item  \label{Low} F. E. Low and P. F. Mende, Ann. Phys. (N.Y.) {\bf 210},
380 (1991).

\item  \label{VDB2} V. Delgado and J. G. Muga, Ann. Phys. (N.Y.) {\bf 248},
122 (1996).

\item  \label{QF} See, for example, C. Itzykson and J. B. Zuber, 
{\it Quantum Field Theory} (McGraw-Hill, New York, 1985).

\item  \label{Aharo} Y. Aharonov and D. Bohm, Phys. Rev. {\bf 122}, 1649
(1961).

\item  \label{Taylor} J. R. Taylor, {\it Scattering Theory: The Quantum
Theory on Nonrelativistic Collisions} (Wiley, New York, 1975).

\item  \label{asymp} N. Bleistein and R. A. Handelsman, {\it Asymptotic
Expansions of Integrals} (Dover, New York, 1986).

\item  \label{Wigner2} M. Hillary, R. F. O'Connell, M. O. Scully, and E. P.
Wigner, Phys. Rep. {\bf 106}, 121 (1984).

\item  \label{Cohen} L. Cohen, J. Math. Phys. {\bf 7}, 781 (1966).

\item  \label{NEW1} Y. Aharonov, J. Oppenheim, S. Popescu, B. Reznik, and 
W. G. Unruh, quant-ph/9709031 (Sep. 14, 1997).

\item  \label{NEW2} See also E. P. Wigner, in {\it Aspects of Quantum 
Theory}, edited by A. Salam and E. P. Wigner (Cambridge University Press, 
London, 1972), p. 237 and references therein.

\item  \label{NEW3} A simple derivation of this result is also given in a 
recent preprint by C. R. Leavens. 

\end{enumerate}

\end{document}